\documentstyle[prc,aps,preprint]{revtex}

\draft
\begin{document}
\tighten

\title{Computing probabilities of very rare events
for Langevin processes: a new method based on 
importance sampling}
\author{O.\ Mazonka$^a$, C.\ Jarzy\' nski$^b$, J.\ B\l ocki$^a$}
\address{(a) Institute for Nuclear Studies, \' Swierk, Poland}
\address{(b) Theoretical Division, Los Alamos National Laboratory, USA}
\date{\today}
\maketitle

\begin{abstract}
Langevin equations are used to model many processes
of physical interest, including low-energy nuclear
collisions.
In this paper we develop a general method for computing
probabilities of very rare events (e.g.\ small fusion 
cross-sections) for processes described by Langevin 
dynamics.
As we demonstrate with numerical examples as well as 
an exactly solvable model, our method can converge to
the desired answer at a rate which is orders of magnitude 
faster than that achieved with direct simulations
of the process in question.
\end{abstract}


\section*{Introduction}
\label{sec:intro}

Langevin methods offer a powerful tool for the
numerical study of low-energy nuclear processes,
such as fission and heavy-ion fusion.
The evolution of nuclei during such events is
typically described using a few collective
degrees of freedom, evolving under both conservative 
and non-conservative forces.
The latter, arising from the coupling of the
collective variables to the intrinsic nucleonic
degrees of freedom, can be modeled by a 
noisy and a dissipative term in a Langevin
description of the collective motion\cite{fro}.
Once such a stochastic equation of motion has
been written down, it is straightforward to 
numerically simulate the process in question,
using a random number generator to supply the
noise.
By repeating the simulation --
with different sequences of random numbers --
one obtains independent realizations
of the process in question, reflecting the
statistical distribution of events 
occurring during an experiment.

The ``direct simulation'' method outlined
above becomes impractical when studying
rare outcomes.
For instance, if we are interested in
computing the very small cross-section for 
the fusion of two heavy nuclei, then 
the vast majority of realizations will end
with the nuclei flying apart, and the number
of simulations required to obtain even a handful
of fusion events may well be prohibitively large.
The purpose of the present paper is to propose a 
general strategy for getting around this problem.

The basic idea which we shall present is 
essentially a dynamical variant of 
{\it importance sampling}, which amounts to
gaining information about one probability
distribution (a ``target'' distribution, $T$), by
choosing randomly from another (a ``sampling''
distribution, $S$) defined on the same space,
and then {\it biasing} -- assigning weights
to -- the points sampled.
The weights are assigned in such a way that the
{\it biased average} of any quantity, over $N$
points drawn independently from $S$, 
and the {\it unbiased average} of that same
quantity over $N$ points drawn from $T$, 
converge to the same value
in the limit of infinitely many samples, 
$N\rightarrow\infty$.
If the biased average converges faster with $N$
than the unbiased one, then 
importance sampling becomes a practical tool
for increasing the efficiency of the numerical 
estimation of the desired average.

In our case, we are interested in Langevin
trajectories describing (for instance) the
collision of two heavy nuclei, with a very small
probability for fusion.
Our {\it target} ensemble, $T$, is then the
statistical distribution of all such trajectories
with, say, a given initial center-of-mass energy
and impact parameter.
The probability of fusion which we wish to 
compute is defined with respect to this ensemble
of trajectories.
Our {\it sampling} ensemble, $S$, is the
distribution of trajectories evolving -- from the
same initial conditions -- under a {\it modified} 
Langevin equation, which by design is far more 
likely to result in fusion.
The scheme which we propose involves running
a number of simulations with the modified equation
of motion (thus obtaining fusion events with good 
statistics), then computing the desired fusion probability 
from this data, by biasing each trajectory.

The organization of this paper is as follows.
In Section \ref{sec:theory}, we derive the 
method which we propose in this paper.
This section begins with a general discussion of 
importance sampling, before focusing specifically
on its application in the context of Langevin 
trajectories.
In Section \ref{sec:pm} we briefly touch on a number
of practical issues associated with the actual
implementation of the method.
Section \ref{sec:numres} presents numerical results,
in which we illustrate the efficiency of the 
method using a schematic model of nuclear collisions.
In Section \ref{sec:exact} we analyze how the
method behaves for the case of an exactly solvable
model, and we conclude in Section \ref{sec:conclusions}.

We should stress at the outset that, while we shall
present our result within the specific context of
computing nuclear fusion probabilities, the method
we propose can in principle be applied quite generally, 
whenever one is interested in rare outcomes of 
processes described by Langevin dynamics.
For a different approach to studying rare events
(developed in the context of chemical transitions),
see the recent work of Chandler and 
collaborators\cite{chandler}.

\section{Theory}
\label{sec:theory}

\subsection{Importance Sampling}

{\it Importance sampling} is based on a very simple
idea, embodied by Eq.\ref{eq:is} below.
Suppose we have some space
($\zeta$-space) on which are defined two normalized 
probability distributions, $p_S(\zeta)$ and $p_T(\zeta)$, 
corresponding to ``sampling'' and ``target''
ensembles, $S$ and $T$.
Supposing furthermore that $p_S(\zeta)>0$ whenever
$p_T(\zeta)>0$, 
let us introduce a {\it biasing function}
\begin{equation}
\label{eq:defbias}
w(\zeta) = {p_T(\zeta)\over p_S(\zeta)},
\end{equation}
defined at all points $\zeta$ for which
$p_S(\zeta)>0$.
Now let $\langle{\cal O}\rangle_S$ and 
$\langle{\cal O}\rangle_T$ denote the averages of
some observable ${\cal O}(\zeta)$ over the two
distributions:
\begin{equation}
\label{eq:defavg}
\langle{\cal O}\rangle_i \equiv \int d\zeta\,
p_i(\zeta)\,{\cal O}(\zeta) \qquad  , \qquad i=S,T.
\end{equation}
If we are interested in computing 
$\langle{\cal O}\rangle_T$, 
then we can do so by repeatedly sampling
from the target ensemble $T$:
\begin{equation}
\label{eq:dir}
\langle{\cal O}\rangle_T =
\lim_{N\rightarrow\infty}
(1/N)\sum_{n=1}^N
{\cal O}(\zeta_n^T),
\end{equation}
where $\zeta_1^T, \zeta_2^T, \cdots$ is a sequence of
points sampled independently from $T$.
However, it follows from Eqs.\ref{eq:defbias}
and \ref{eq:defavg} above that we can equally
well express the desired average as:
\begin{equation}
\label{eq:is}
\langle{\cal O}\rangle_T = 
\langle w{\cal O}\rangle_S
= \lim_{N\rightarrow\infty}
(1/N)\sum_{n=1}^N 
w(\zeta_n^S)\,{\cal O}(\zeta_n^S),
\end{equation}
where $\zeta_1^S, \zeta_2^S, \cdots$ is a sequence of
points sampled independently from $S$.
Thus, provided we can compute $w(\zeta)$ and
${\cal O}(\zeta)$ for any $\zeta$, Eq.\ref{eq:is}
gives us a prescription for determining the
average of ${\cal O}$ over the target ensemble
$T$, {\it using points drawn from the sampling 
ensemble} $S$.
This prescription becomes a useful tool
if a sampling distribution can be chosen
for which the rate of convergence
with the number of samples ($N$) is faster
when using Eq.\ref{eq:is}, than when sampling
directly from $T$ (Eq.\ref{eq:dir}).

Let us now consider how we might apply
importance sampling to the problem which interests
us, namely, computing the probability of fusion
for two colliding nuclei.
For simplicity of presentation, we assume 
for the moment that the
collision process is described by the evolution of
a single collective degree of freedom (e.g.\ a
distance coordinate between the two nuclei),
obeying a Langevin equation of the form
\begin{equation}
\label{eq:original}
{dx\over dt} = v_0(x) + \hat\xi(t).
\end{equation}
Here, $x$ is the collective variable, 
$v_0(x)$ is a ``drift'' term which embodies the
deterministic forces -- both conservative and
dissipative -- acting on the collective degree of
freedom, and $\hat\xi(t)$ is a stochastic, white noise
term:
\begin{equation}
\label{eq:xi}
\langle \hat\xi(t) \hat\xi(t+s) \rangle = D \delta(s),
\end{equation}
where $\langle\cdots\rangle$ denotes an average
over realizations of $\hat\xi(t)$, and $D>0$ is a
diffusion constant.
[More generally, the number of variables required to
specify the instantaneous state of the system will
be greater than one:
$x\rightarrow\vec x=(x_1,\cdots,x_d)$.
In the case of overdamped motion, this vector
specifies the instantaneous configuration of the
colliding nuclei.
For evolution in which inertial effects are
important, the vector $\vec x$ will include both
configurational variables and their associated
momenta.
See Section \ref{sec:pm} for elaboration of these
and other points.]

Let us assume initial conditions corresponding to
a particular value $x^0$ for the collective variable,
and suppose we are interested in the evolution of
the colliding nuclei over a time interval
$0\le t\le\tau$.
Then a single realization of this process is described
by a trajectory $x(t)$, $0\le t\le\tau$, satisfying
$x(0)=x^0$, which obeys Eq.\ref{eq:original} for a
given realization of the noise term $\hat\xi(t)$.
Therefore, the statistical ensemble of realizations
of $\hat\xi(t)$ -- together with Eq.\ref{eq:original}
and the initial condition $x(0)=x^0$ --
defines a statistical ensemble of trajectories $x(t)$.
By simulating such a process numerically --
using a random number generator to provide the
noise term -- we are, effectively, sampling 
randomly from this ensemble.
Let $T$ (``target'') denote the 
ensemble of trajectories thus defined:
\begin{equation}
\label{eq:t}
T \equiv \Bigl\{ x(t) \,\Bigl\vert\,
\dot x = v_0 + \hat\xi \, , \,
x(0) = x^0 \Bigr\},
\end{equation}
with the time interval $0\le t\le\tau$ left
implicit, here and henceforth. 

Given the above Langevin equation (Eq.\ref{eq:original}),
along with the initial condition $x^0$, we are
interested in computing the probability of fusion.
Here we will take ``fusion'' to mean simply that the
final point of the trajectory, $x(\tau)$, falls within
a given region $R$ along the $x$-axis:
\begin{eqnarray}
x(\tau) \in R \quad &\rightarrow& \quad {\rm fusion}  \\
x(\tau) \notin R \quad &\rightarrow& \quad {\rm no\,\,fusion}.
\end{eqnarray}
Now introduce a functional $\Theta[x(t)]$ which
is equal to 1 if the trajectory yields a 
fusion event, and 0 otherwise.\footnote{
Of course, in our case, the functional 
$\Theta[x(t)]$ is really just a function of the
final point, $x(\tau)$. 
More generally, however, we could have
introduced a criterion for ``fusion''
which depends on the entire trajectory,
in which case $\Theta[x(t)]$ would be a genuine
function{\it al}.}
Then the probability of fusion, $P_{\rm fus}$, is 
just the average of this functional over the 
ensemble $T$, defined by Eq.\ref{eq:t} above:
\begin{equation}
P_{\rm fus} = \langle \Theta \rangle_T.
\end{equation}

We can compute $P_{\rm fus}$ by sampling randomly
from the ensemble $T$ (i.e.\ repeatedly simulating
trajectories evolving under Eq.\ref{eq:original}),
and counting the number of fusion events:
\begin{equation}
\label{eq:direct}
P_{\rm fus} \cong P_{\rm fus}^{(N)} = 
{1\over N} \sum_{n=1}^N \Theta[x_n^T(t)],
\end{equation}
where $x_n^T(t)$ is the $n$'th of $N$ independent
simulations of the process, all launched from $x^0$.
However, since the number of simulations needed to
compute $P_{\rm fus}$ to a desired accuracy grows
as the inverse of the fusion probability itself
(see Eq.\ref{eq:nzero}), this ``direct sampling'' method
becomes impractical for very small values of $P_{\rm fus}$.
Let us therefore consider a modified version of
Eq.\ref{eq:original}:
\begin{equation}
\label{eq:modified}
{dx\over dt} = v_0(x) + \Delta v(x) + \hat\xi(t),
\end{equation}
where the modification consists of adding an
extra drift term, $\Delta v$, chosen to greatly 
increase the probability of simulating a fusion event.
Let $S$ (for ``sampling'') denote the statistical
ensemble of trajectories $x(t)$ corresponding
to {\it this} equation of motion, with the
same initial conditions as above:
\begin{equation}
\label{eq:s}
S \equiv \Bigl\{ x(t) \,\Bigl\vert\,
\dot x = v_0 + \Delta v + \hat\xi \, , \,
x(0) = x^0 \Bigr\}.
\end{equation}
Now, for a given trajectory $x(t)$ satisfying
$x(0)=x^0$, let $p_T[x(t)]$ 
denote the probability density that we will obtain
this particular trajectory, under the original
Langevin equation (Eq.\ref{eq:original});
and let $p_S[x(t)]$ denote
the probability density for obtaining this trajectory
under the modified equation (Eq.\ref{eq:modified}).
Finally, let $w$ denote the ratio of these two
densities\footnote{
Note that while the values of
$p_S[x(t)]$ and $p_T[x(t)]$ depend on the {\it measure}
chosen on path space, the ratio $w[x(t)]$ is
independent of that measure.}:
\begin{equation}
w[x(t)] \equiv {p_T[x(t)] \over p_S[x(t)]}.
\end{equation}
Then, by Eq.\ref{eq:is}, we can express the
fusion probability $P_{\rm fus}$ (defined with
respect to the {\it original} Langevin equation) as:
\begin{equation}
\label{eq:isfus}
P_{\rm fus} = 
\langle w\Theta\rangle_S = 
\lim_{N\rightarrow\infty} {1\over N}
\sum_{n=1}^N w[x_n^S(t)]\cdot\Theta[x_n^S(t)],
\end{equation}
where $x_n^S(t)$ is the $n$'th
of $N$ independent realizations obeying the
{\it modified} Langevin equation.
This means that, if we know how to compute $w$
for any trajectory $x(t)$, then we can estimate
$P_{\rm fus}$ by
running $N$ simulations with the
modified Langevin equation, then adding together
the weights $w$ for all those trajectories which
lead to fusion, and dividing by the total number
of simulations, $N$. 
In the limit $N\rightarrow\infty$, this estimate
converges to the correct value of $P_{\rm fus}$.

In the following section we argue that there indeed
exists an explicit, easily computable expression
for $w[x(t)]$, which renders practical the
importance sampling strategy outlined above.

\subsection{Statistical distributions of Langevin trajectories}
\label{subsec:statlang}

The original and modified Langevin equations,
Eqs.\ref{eq:original} and \ref{eq:modified}, can 
be represented by the generic equation
\begin{equation}
\label{eq:lang}
{dx\over dt} = v(x) + \hat\xi(t),
\end{equation}
where $v=v_0$ in one case, and $v=v_0+\Delta v$ in
the other.
As before, given some initial conditions $x(0)=x^0$, 
let $x(t)$ denote the 
trajectory evolving from those initial conditions,
for a particular realization of the noise term.
We are interested in the probability density $p[x(t)]$ 
for obtaining a particular trajectory $x(t)$.
To introduce a measure on the space of all possible 
trajectories, we discretize the trajectory.
Thus, let $(x^0,x^1,\cdots,x^M)$ denote the set
of points specifying the state of the system, 
after time intervals $\delta t = \tau/M$:
\begin{equation}
x^m = x(m\,\delta t)\qquad,\qquad
m=0,\cdots, M.
\end{equation}
We can then ask for the probability density
$p(x^1,\cdots,x^M\vert x^0)$ that the trajectory
passes through the sequence of points
$(x^1,x^2,\cdots,x^M)$ at times
$\delta t, 2\delta t,\cdots,\tau$,
given the initial point $x^0$.
Note that this is a probability distribution in
$M$-dimensional $(x^1,\cdots,x^M)$-space, with
$x^0$ acting as a parameter of the distribution.
Let us introduce the notation
${\bf X} = (x^0,x^1,\cdots,x^M) = (x^0,{\bf Y})$.
Then an explicit expression for $p$ is given 
by\cite{polymer}:
\begin{mathletters}
\label{eq:proba}
\begin{equation}
\label{eq:p}
p({\bf Y}\vert x^0) = 
(2\pi D\delta t)^{-M/2}
\exp -A({\bf X}),
\end{equation}
where 
\begin{equation}
\label{eq:a}
A({\bf X}) \equiv {1\over 2D\delta t}
\sum_{m=0}^{M-1} \Bigl[x^{m+1}-x^m-v(x^m)\delta t\Bigr]^2.
\end{equation}
\end{mathletters}
Eq.\ref{eq:p} is strictly speaking valid only in
the limit $M\rightarrow\infty$ (with $\tau$ fixed),
although it constitutes a good approximation if 
$\sqrt{D\delta t}$ is small in comparison with the
length scale set by variations in $v(x)$.

Eq.\ref{eq:proba} is a general expression for the
probability density of obtaining a particular
discretized trajectory ${\bf X}$, launched from $x^0$.
Now let $p_T({\bf Y}\vert x^0)$ and $p_S({\bf Y}\vert x^0)$
denote this probability density, for the specific
Langevin processes described by Eqs.\ref{eq:original}
and \ref{eq:modified}, respectively.
Then the ratio between these two probability
densities is given by:
\begin{mathletters}
\label{eq:weight}
\begin{eqnarray}
w({\bf X}) &\equiv&
{p_T({\bf Y}\vert x^0) \over p_S({\bf Y}\vert x^0)}
= \exp -\Delta A({\bf X}) \\
\Delta A &\equiv& A_T - A_S,
\end{eqnarray}
\end{mathletters}
where $A_T$ and $A_S$ are computed with Eq.\ref{eq:a},
using $v=v_0$ and $v=v_0+\Delta v$, respectively.
If we now explicitly write out the expression for
$\Delta A$ in terms of $(x^0,x^1,\cdots,x^M)$, and 
consider the limit $M\rightarrow\infty$ (with
$\tau$ fixed), then we arrive at the following
result for our biasing function, expressed
in terms of an integral over a 
trajectory $x(t)$ rather than a sum over 
points along the trajectory:
\begin{mathletters}
\label{eq:wtraj}
\begin{eqnarray}
w[x(t)] &=& \exp -\Delta A \\
\label{eq:da}
\Delta A[x(t)] &=& {1\over D} \int_0^\tau dt
\Biggl( {dx\over dt} - v_0 - {1\over 2} \Delta v \Biggr)
\Delta v.
\end{eqnarray}
\end{mathletters}
Here, $dx/dt$, $v_0$ and $\Delta v$ are evaluated 
along the trajectory $x(t)$.

\subsection{Computing probabilities of rare events}

Combining Eqs.\ref{eq:isfus} and
\ref{eq:wtraj}, we now have an expression which
allows us, in principle, to compute the 
probability for fusion -- defined with respect to
the {\it original} equation of motion (Eq.\ref{eq:original}) 
-- by running independent simulations with the
{\it modified} equation of motion (Eq.\ref{eq:modified}):
\begin{equation}
\label{eq:pfusis}
P_{\rm fus} = \lim_{N\rightarrow\infty}
{1\over N}
\sum_{n=1}^N \Theta[x_n^S(t)]\cdot
\exp -\Delta A[x_n^S(t)].
\end{equation}
Here, $x_n^S(t)$ is the trajectory
generated during the $n$'th simulation, using
the modified Langevin equation;
$\Delta A$ is computed for each trajectory\footnote{
While Eq.\ref{eq:da} gives a well-defined expression
for $\Delta A$, in practice it is more convenient
to compute $\Delta A$ using the method described in
Section \ref{sec:pm} below; see Eqs.\ref{eq:newda}
to \ref{eq:daytau}.}; and $\Theta$ is, as before, 
equal to one or zero, depending on whether or not 
fusion occurred.

For a given set of original and modified
Langevin equations, and for a large number $N$
of trajectories simulated under the modified
equations, we thus have the following 
estimator for $P_{\rm fus}$:
\begin{equation}
\label{eq:estimator}
P_{\rm fus} \cong P_{\rm fus}^{(N)} =
{1\over N}
\sum_{n=1}^N \Theta_n
\exp -\Delta A_n ,
\end{equation}
where $\Theta_n\equiv\Theta[x_n^S(t)]$, and 
$\Delta A_n\equiv\Delta A[x_n^S(t)]$.
The estimator $P_{\rm fus}^{(N)}$
converges to the true value $P_{\rm fus}$ in the
limit of infinitely many trajectories:
$\lim_{N\rightarrow\infty} P_{\rm fus}^{(N)} 
= P_{\rm fus}$.

\subsection{Efficiency analysis}
\label{subsec:eff}

Having derived an estimator for $P_{\rm fus}$
based on the idea of importance sampling,
we now consider the question of efficiency.
In particular, we establish a specific measure
of ``how much we gain'' by using importance sampling,
with a given choice of $\Delta v(x)$.

The validity of Eq.\ref{eq:pfusis} does not
depend on the form of $\Delta v(x)$.
Therefore, for any additional drift term
$\Delta v$, there will be some threshold
value $N_{\Delta v}^*$ such that 
$P_{\rm fus}^{(N)}$ provides a ``good'' estimate 
of $P_{\rm fus}$ for $N\ge N_{\Delta v}^*$.
That is, $N_{\Delta v}^*$ is the number
of trajectories which we need to simulate
(with the modified Langevin equation), in
order to determine $P_{\rm fus}$ to some
desired accuracy, using the method outlined
above.
Of course, $N_{\Delta v}^*$ can depend strongly
on the form of $\Delta v(x)$.
We can thus compare the efficiency of estimating
$P_{\rm fus}$, for different drift terms.
In particular -- since the special case
$\Delta v=0$ is equivalent to computing 
$P_{\rm fus}$ using the original Langevin
equation -- let us define the {\it efficiency
gain}, $E_{\Delta v}^G$, associated with a
given $\Delta v(x)$, as follows:
\begin{equation}
E_{\Delta v}^G \equiv
{N_0^* \over N_{\Delta v}^*}.
\end{equation}
The numerator is just the number of trajectories
needed to accurately estimate $P_{\rm fus}$
by running simulations with the original
Langevin equation ($\Delta v=0$);
the denominator is the number needed using
Eq.\ref{eq:estimator}, for a given 
$\Delta v(x)$.
Thus\footnote{
We are assuming that simulating a single trajectory
with the modified Langevin equation takes as
much time as simulating one with the original
Langevin equation.}
, $E_{\Delta v}^G$
is the factor by which we reduce the 
computational effort, by making use of
importance sampling -- again, for a given
$\Delta v(x)$.

Let us derive an expression for $E_{\Delta v}^G$
in terms of quantities extracted
directly from numerical simulations.
For a given additional drift term $\Delta v$,
let us define
\begin{equation}
f[x(t)] \equiv 
w[x(t)]\cdot\Theta[x(t)] 
= \Theta\,\exp -\Delta A.
\end{equation}
Our method of computing $P_{\rm fus}$ then
amounts to computing the average of $f$, 
by sampling trajectories $x(t)$ from the 
ensemble $S$:
$\langle f\rangle_S = P_{\rm fus}\cong 
P_{\rm fus}^{(N)} = (1/N)\sum_n f_n$,
where $f_n=f[x_n^S(t)]$.
The statistical error in our result --
the expected amount by which 
$P_{\rm fus}^{(N)}$ differs from $P_{\rm fus}$ --
is given by the usual formula for the 
standard deviation of the mean:
\begin{equation}
\sigma_{P_{\rm fus}}^{(S)} = {\sigma_f\over\sqrt{N}} ,
\end{equation}
where $\sigma_f^2$ is the variance of the 
quantity $f$ over the sampling ensemble,
\begin{equation}
\sigma_f^2 = \langle f^2\rangle_S - \langle f\rangle_S^2.
\end{equation}

If we want to compute $P_{\rm fus}$ to a desired
relative accuracy $r$ --
in the sense that we want the ratio 
$\sigma_{P_{\rm fus}}/P_{\rm fus}$
(expected error / desired average)
to fall below $r$ --
then we get the following expression for the
minimum number of trajectories needed:
\begin{equation}
\label{eq:ndv}
N_{\Delta v}^* = 
{\sigma_f^2 \over r^2 P_{\rm fus}^2}
= {1\over r^2} \cdot 
{ \langle f^2\rangle_S - \langle f\rangle_S^2 \over
\langle f\rangle_S^2 }.
\end{equation}
In other words, if we simulate more than 
this many trajectories, then we can expect
the statistical error in our final estimate of $P_{\rm fus}$
to be no greater than $r$ times $P_{\rm fus}$.

In the case of direct sampling from the target
ensemble $T$, we estimate $P_{\rm fus}$
by computing the average of $\Theta$.
The expected statistical error in this average
is just
\begin{equation}
\sigma_{P_{\rm fus}}^{(T)} = 
\Biggl[
{\langle\Theta^2\rangle_T - \langle\Theta\rangle_T^2
\over N} \Biggr]^{1/2} = 
\Biggl[
{P_{\rm fus} (1-P_{\rm fus}) \over N} \Biggr]^{1/2},
\end{equation}
making use of the fact that $\Theta^2=\Theta$.
By setting this expected statistical error equal
to $rP_{\rm fus}$, we get
\begin{equation}
\label{eq:nzero}
N_0^* = {1\over r^2}\cdot
{1-P_{\rm fus} \over P_{\rm fus}} \cong
{1\over r^2 P_{\rm fus}},
\end{equation}
for $P_{\rm fus}\ll 1$.
(If we simulate this many trajectories,
then the expected number of fusion events is
$1/r^2$, and the expected statistical error in
this number is $1/r$.)

Combining Eqs.\ref{eq:ndv} and \ref{eq:nzero}, we
get the following result for the efficiency gain
of our importance sampling
method, for a particular choice of $\Delta v(x)$:
\begin{equation}
\label{eq:edv}
E_{\Delta v}^G = { N_0^* \over N_{\Delta v}^* }
= {P_{\rm fus} (1-P_{\rm fus}) \over
\sigma_f} \cong
{ P_{\rm fus} \over
\sigma_f}  =
{ \langle f\rangle_S \over
\langle f^2\rangle_S - \langle f\rangle_S^2 }.
\end{equation}

Eq.\ref{eq:edv} gives the efficiency gain of
importance sampling, with a particular choice of
$\Delta v$, in terms of averages which can be estimated 
from simulations performed under the modified Langevin
equation alone (i.e.\ sampling only from $S$, not $T$).
An expression for $E_{\Delta v}^G$ in terms
of averages estimated using both the original and
modified equations of motion, is:
\begin{equation}
\label{eq:edv2}
E_{\Delta v}^G = 
\Biggl[
{\sigma_{P_{\rm fus}}^{(T)} \over
\sigma_{P_{\rm fus}}^{(S)}}
\Biggr]^2 = 
{ \langle \Theta\rangle_T - \langle \Theta\rangle_T^2 \over
  \langle f^2\rangle_S - \langle f\rangle_S^2 }.
\end{equation}

We will use these results in Sections \ref{sec:numres}
and \ref{sec:exact} below, to compute the 
efficiency gain of our importance sampling method for
particular examples.

\section{Practical Matters}
\label{sec:pm}

In this section, we discuss a number of practical issues
related to the actual implementation of our method.

$\bullet$ 
At the end of Section \ref{subsec:statlang}, 
we obtained an expression for $\Delta A$ as a functional 
of the trajectory $x(t)$.
Since $x(t)$ satisfies Eq.\ref{eq:modified}, we can
rewrite Eq.\ref{eq:da} as:
\begin{equation}
\label{eq:newda}
\Delta A = {1\over 2D} \int_0^\tau dt\,
(2\hat\xi + \Delta v) \Delta v.
\end{equation}
This expression for $\Delta A$ lends itself to a 
convenient implementation of our method, as follows.
When simulating a given trajectory $x(t)$ evolving
under Eq.\ref{eq:modified}, we simultaneously
integrate the following equation of motion for
a new variable $y(t)$, satisfying the initial
condition $y(0)=0$:
\begin{equation}
\label{eq:dydt}
{dy\over dt} = 
{\Delta v\over 2D} (2\hat\xi + \Delta v), 
\end{equation}
for the same realization of the noise term
$\hat\xi(t)$.
(Note that this equation is coupled to 
the equation of motion for $x$, since $\Delta v$
in general depends on $x$.)
Eq.\ref{eq:newda} then implies that
\begin{equation}
\label{eq:daytau}
\Delta A = y(\tau).
\end{equation}
Thus, at the end of the simulation, we use
$x(\tau)$ to determine whether or not fusion has
occurred, and if so, then we take $\Delta A = y(\tau)$
when assigning the bias $e^{-\Delta A}$ to this
event.

$\bullet$ 
Often (see for instance Section \ref{sec:numres} below), 
the evolution of our system is such that, once a trajectory 
$x(t)$ enters the region $R$ which defines fusion, its chance
for subsequently escaping that region is negligible:
$R$ effectively possesses an absorbing boundary.  
If this is true for both the original and modified
evolution, it becomes convenient to define
$\Delta v$ to be zero everywhere within $R$.
Then, if a trajectory $x(t)$ (evolving under the
modified Langevin equation) crosses into $R$ at some
time $\tau^\prime < \tau$, we can stop the simulation
at that point in time, and take
$\Theta = 1$, $\Delta A = y(\tau^\prime)$.
This saves time, by eliminating the need to continue
with the simulation.

$\bullet$
We have, to this point, assumed that the stochastic
noise $\hat\xi(t)$ is independent of $x$. 
That is, $D=$const.
More generally, we might have a diffusion coefficient
which depends on the instantaneous configuration of
the system: $D=D(x)$, i.e.
\begin{equation}
\langle \hat\xi(t)\hat\xi(t+s) \rangle_{x(t)=x}
= D(x)\delta(s).
\end{equation}
In this case Eq.\ref{eq:proba} becomes
\begin{eqnarray}
\label{eq:dofx}
p({\bf Y}|x^0) &=&
\Biggl\{
\prod_{m=0}^{M-1}
\Bigl[2\pi D(x^m)\delta t\Bigr]^{-1/2}
\Biggr\}
\exp{-A({\bf X})} \\
A({\bf X}) &=& {1\over 2 \delta t}
\sum_{m=0}^{M-1}{1\over D(x^m)}\Bigl[x^{m+1}-x^m-v(x^m)
\delta t\Bigr]^2\qquad .
\end{eqnarray}
Eq.\ref{eq:weight} remains unchanged (since the 
factor multiplying $e^{-A}$ in Eq.\ref{eq:dofx}
is the same for $p_S$ and $p_T$);
and Eq.\ref{eq:wtraj} changes only in that $1/D$ is
brought inside the integral in Eq.\ref{eq:da},
with $D$ evaluated along the trajectory $x(t)$.
When implementing the method using the additional
variable $y(t)$, the only difference is that
$D$ is evaluated along $x(t)$ rather
than being a constant, in Eq.\ref{eq:dydt}.

$\bullet$
Let us now drop the assumption that the system
evolves in one dimension.
The state of the system is now described by a 
vector $\vec x = (x_1,\cdots,x_d)$, evolving under a
set of coupled Langevin equations,
\begin{equation}
{d x_i\over dt} = v_i(\vec x) + \hat\xi_i(t)
\qquad,\qquad i=1,\cdots,d,
\end{equation}
where $\vec v = \vec v_0$ (original equations
of motion) or $\vec v = \vec v_0 + \Delta\vec v$ 
(modified equations of motion).
The diffusion coefficient $D$ becomes a symmetric
matrix whose elements reflect the correlations between 
the different components of the stochastic force:
\begin{equation}
\langle \hat\xi_i(t) \hat\xi_j(t+s) \rangle = D_{ij}\delta(s).
\end{equation}
[For simplicity, we assume that this matrix is a
constant.
The generalization to $D_{ij} = D_{ij}(\vec x)$ is as
straightforward as in the one-dimensional case.]

The simplest case of multi-dimensional evolution
occurs when the components $\hat\xi_i$ are mutually
uncorrelated.
Then $D$ is a diagonal matrix, and the generalization
of Eq.\ref{eq:dydt} is given by:
\begin{equation}
\label{eq:diag}
{dy\over dt} =
\sum_{i,D_{ii}\ne 0} {\Delta v_i\over 2D_{ii}}
(2\hat\xi_i + \Delta v_i).
\end{equation}
The sum is taken over values of $i$ for which 
$D_{ii}\ne 0$, corresponding to those directions along
which there is a non-zero stochastic force.
Along those directions for which $D_{ii}=0$,
we must have $\Delta v_i=0$.\footnote{
To see this, note that if $D_{ii}=0$ for a particular 
value of $i$, then the equation of motion describing 
the evolution of $x_i$ is deterministic
($\hat\xi_i=0$).
If we now modify that equation by adding an
additional non-zero term $\Delta v_i$, then any trajectory
${\bf x}(t)$ obeying the modified equations of motion
will not be a solution of the original equations -- for
any realization of the noise term $\hat\xi(t)$ -- and
vice-versa (unless $\Delta v_i$ happens to be zero
exactly along the trajectory).
This violates the condition stated before 
Eq.\ref{eq:defbias}:
in order for the importance sampling to be valid,
our modified equations of motion must be capable
of generating any trajectory which might be generated
by the original equations of motion.}

If $D$ is not diagonal, then Eq.\ref{eq:diag} 
generalizes to the following evolution equation for
$y$:
\begin{equation}
\label{eq:nondiag}
{dy\over dt} = {1\over 2}
(2{\vec\xi} + {\Delta\vec v})^T D^{-1} {\Delta\vec v}.
\end{equation}
Eq.\ref{eq:nondiag} implicitly assumes that $D$ is
invertible, i.e.\ ${\rm det}(D)\ne 0$.
If this is not the case, then -- first -- we must
make sure that the projection of ${\Delta\vec v}$
onto the subspace spanned by the null eigenvectors
of $D$ is zero.
(See the comments following Eq.\ref{eq:diag}.)
Assuming this condition is satisfied, we can view
Eq.\ref{eq:nondiag} as pertaining only to the 
subspace spanned by the non-zero eigenvectors of $D$.

The results discussed in this generalization from
one-dimensional to multi-dimensional evolution are based 
on the following generalization of Eq.\ref{eq:proba}:
\begin{mathletters}
\begin{eqnarray}
p({\bf Y}\vert x^0) &=&
\Bigl[
(2\pi\delta t)^{d^\prime}\,
{\rm det}(D)\Bigr]^{-M/2} \exp -A({\bf X})\\
A({\bf X}) &=& {1\over 2\delta t}
\sum_{m=0}^{M-1}
[\vec x^{m+1}-\vec x^m-\vec v(\vec x^m)\delta t]^T
D^{-1}
[\vec x^{m+1}-\vec x^m-\vec v(\vec x^m)\delta t].
\end{eqnarray}
\end{mathletters}
As with Eq.\ref{eq:nondiag} above, these equations pertain
to the $d^\prime(\le d)$-dimensional subspace spanned
by the non-zero eigenvectors of $D$.

There is an example of multi-dimensional
evolution worthy of particular mention.
This is the case in which inertial effects are present, i.e.\
the evolution of the system is not overdamped.
The evolution then occurs in the {\it phase space}
of the system, and the equations of motion for 
${\vec x}=({\vec q},{\vec p})$ are typically of the form:
\begin{mathletters}
\label{eq:inertial}
\begin{eqnarray}
\label{eq:confeq}
{d{\vec q}\over dt} &=& I^{-1} {\vec p} \\
\label{eq:momeq}
{d{\vec p}\over dt} &=& {\vec F} - \Gamma I^{-1} {\vec p} 
+ {\vec\xi}.
\end{eqnarray}
\end{mathletters}
Here, ${\vec q}$ is a vector of variables specifying
the configuration of the system;
${\vec p}$ is the vector of associated momenta;
$I$ is an inertia tensor;
${\vec F}$ is the vector of conservative forces acting
on the system;
$\Gamma$ is a friction tensor;
and ${\vec\xi}$ is the vector of stochastic forces,
whose associated diffusion tensor $D$ is related to
$\Gamma$ by a fluctuation-dissipation relation.
(Typically, $I$, ${\vec F}$, $\Gamma$, and $D$ are all
functions of ${\vec q}$.)
In this case, the equations of motion for $\vec q$
are deterministic (Eq.\ref{eq:confeq}), therefore
any additional drift terms 
must appear only in the {\it momentum} equations
(Eq.\ref{eq:momeq}), as an additional force
$\Delta{\vec F}$.
The equation of motion for $y(t)$, Eq.\ref{eq:nondiag},
then pertains only to momentum space 
(the ${\vec p}$-subspace of phase space).

$\bullet$
Finally, it is often the situation that the 
dissipative and stochastic forces acting on the 
collective degrees of freedom depend on the 
``temperature'' of the bi-nuclear system.
This is another way of saying that these forces,
at time $t$, depend on the total amount of collective 
energy which has been dissipated up to that time.
Since the energy dissipated, as a function of time,
will differ from one realization to the next,
it seems we are faced with ``memory-dependent''
forces, i.e.\ forces which, at time $t$, depend
not only on the instantaneous state of the system,
but also on its {\it history}, up to time $t$.
An easy way to deal with this situation is simply
to expand our list of variables $\vec x$ to 
include a new member, $x_{d+1}=E_{\rm diss}$,  
denoting the collective energy dissipated.
This new variable is initialized at zero, and evolves
under an evolution equation which depends on the
model used to describe the colliding nuclei.
(For instance,
\begin{equation}
{dE_{\rm diss}\over dt} = 
(I^{-1}\vec p)^T
[\Gamma I^{-1}\vec p -\vec\xi]
\end{equation}
in the case of collective evolution described by
Eq.\ref{eq:inertial}.)

With the addition of this new variable, 
i.e.\ $\vec x\rightarrow (x_1,\cdots,x_d,x_{d+1})$, 
we now again have a set of (coupled) Langevin
equations, in which the drift and stochastic terms
depend only on the instantaneous state of the
system, $\vec x$.
We can thus apply the method proposed in this
paper, without further modification.

\section{Numerical Results}
\label{sec:numres}

In this Section we describe numerical experiments
which we have carried out to test our method, using a 
simplified model of heavy ion collisions
introduced by \' Swi\c atecki\cite{wladek}.
This model was previously studied by Aguiar {\it et al} 
in 1990\cite{aguiar}, using Langevin simulations.
For our example, we considered the collision of two 
$^{100}$Zr nuclei. 
In this mass-symmetric case - for this simple model - 
the shape of the system is defined by two equal spheres 
connected by a cylinder. 
There are two macroscopic (``collective'')
variables parametrizing the shape: (1) the {\it relative
distance} $\rho$ between the sphere centers, which is
the distance $s$ divided by the sum of radii of the 
two spheres: $\rho = s/2R$; and (2) the
{\it window opening} $\alpha$, which is the square
of the ratio of the cylinder radius to the radius of
the sphere: $\alpha = (r_{cyl}/R)^2$.

After some approximations for the potential, dissipation
and kinetic energy terms, one obtains the following
coupled differential equations for the time evolution of
the system (see Ref.\cite{wladek} for details):
\begin{mathletters}
\label{eq:schem_orig}
\begin{eqnarray}
\label{eq:schem_orig_a}
\mu {d^2\sigma\over d\tau^2} + 
\nu^2 {d\sigma\over d\tau} + \nu - X 
&=& \hat\xi_1 \\
{d\nu\over d\tau} - 
{2\nu+3\nu^2-\sigma\over 4\nu(\sigma+\nu^2)}
&=& \hat\xi_2.
\end{eqnarray}
\end{mathletters}
(While Eq.\ref{eq:schem_orig_a} has been written here
as a second-order stochastic differential equation,
in practice we convert it to two first-order
equations -- one deterministic, one stochastic --
by introducing the variable $p_\sigma = \mu\,d\sigma/d\tau$.)
Here, the collective coordinates $\rho$ and $\alpha$
are represented by the variables $\nu = \sqrt{\alpha}$ 
and $\sigma = \rho^2-1$;
the constant $\mu$ is a reduced mass, $\tau$ is a reduced
time, $X$ is a constant
conservative force, and $\hat\xi_1$ and $\hat\xi_2$ are 
stochastic forces
(Gaussian white noise), related to the dissipative terms
by a fluctuation-dissipation relation.
(For the $^{100}$Zr$+^{100}$Zr collision, we have
$\mu\cong 0.176$ and $X\cong 0.677$.)
The evolution of the colliding nuclei is then represented
by a Langevin trajectory in $(\sigma,\nu)$-space.
Fig.\ref{fig:one} depicts 30 such trajectories,
all starting from a configuration of two touching spheres
($\sigma = 0, \nu = 0 $),
with a center-of-mass energy equal to 0.8 MeV above
the interaction barrier.
This energy is about 2.5 MeV below the ``extra push''
energy, so most of the trajectories (28 of them) lead 
to reseparation of the system (fission), and only two
trajectories lead to a compound nucleus (fusion).

From Fig.\ref{fig:one} we have the following picture of 
the physical process occuring, in the context of this
simplified model:
first the window opening between the two nuclei
grows rapidly; then around a saddle point, at
$(\sigma,\nu)\sim(0.0,0.6)$, the combination of
deterministic and stochastic forces determines the
ultimate fate of the nuclei, either fusion or 
reseparation; and finally the system evolves toward
its destiny, with $\sigma$ decreasing in the case 
of fusion, or increasing with reseparation.
This suggests that, if we are to add an extra
drift term to increase the likelihood of fusion, then
it would be best to localize such a term in the
vicinity of the saddle point.
We will think of such a term as a {\it force},
pushing the system toward fusion.
We have chosen an additional force along the
(negative) $\sigma$ direction, whose strength is
a Gaussian function of $(\sigma,\nu)$, with a peak
at $(0.0,0.6)$.
This leads to the following {\it modified} Langevin
equations of motion:
\begin{mathletters}
\label{eq:schem_mod}
\begin{eqnarray}
\mu {d^2\sigma\over d\tau^2} + 
\nu^2 {d\sigma\over d\tau} + \nu - X 
&=& \hat\xi_1 - \Lambda \exp\Biggl\{
-{\sigma^2+(\nu-0.6)^2\over 0.02}
\Biggr\}
\\
{d\nu\over d\tau} - 
{2\nu+3\nu^2-\sigma\over 4\nu(\sigma+\nu^2)}
&=& \hat\xi_2\quad,
\end{eqnarray}
\end{mathletters}
with $\Lambda$ an adjustable parameter.
Fig.\ref{fig:two} schematically shows the region around
the saddle point, where the additional force pushing the
system toward fusion is localized.
Two deterministic trajectories (evolving under the
original equations of motion, but without the stochastic
terms) are also shown, to guide the eye.
One of these was launched with an energy of 1 MeV
above the barrier (leading to reseparation),
the other one at 5 MeV above the barrier (leading
to fusion).

To compare our importance sampling method to
direct simulation of the original process, 
we first chose to compute the probability of fusion 
for trajectories starting with an energy 0.2 MeV above 
the barrier.
This probability is on the order of $10^{-3}$, 
considerably less than that for the case shown in 
Fig.\ref{fig:one} (0.8 MeV above the barrier).
We ran $10^4$ independent trajectories under both
the original and the modified Langevin equations,
Eqs.\ref{eq:schem_orig} and \ref{eq:schem_mod},
respectively,
and kept a running tally of the probablity of 
fusion, as computed by the two methods
(Eq.\ref{eq:direct} for the case of direct
simulation, Eq.\ref{eq:estimator} for importance
sampling).
We took the strength of the additional force to be
$\Lambda=0.3\mu$ in these simulations.
Fig.\ref{fig:three} illustrates the difference
between the rates of convergence of the two methods.
The plots show the estimates $P_{\rm fus}^{(N)}$ as
computed by each method, as a function of number of
trajectories simulated, $N$.
It is clear that the importance sampling (broken line)
converges much faster than direct simulation (solid
line):  
after about 5000 trajectories, the former has 
converged very close to its asymptotic value, 
whereas the latter is still ``jumping around''
after 10000 trajectories.

The sawtooth pattern exhibited by the direct
simulation estimate is typical of the situation
in which rare events contribute disproportionally
to an ensemble average:
in this particular case, only 15 of the 
$10000$ ``direct'' trajectories (simulated under
Eq.\ref{eq:schem_orig}) lead to fusion; each
of these events causes a sudden jump in the 
$P_{\rm fus}^{(N)}$, followed by a gradual
decline as non-fusion events accumulate.
By contrast, under the modified Langevin equations,
Eq.\ref{eq:schem_mod}, about $25\%$ of the 
$10000$ trajectories
lead to fusion, resulting in smoother and faster 
convergence.

In Figs.\ref{fig:four} and \ref{fig:five} we show
excitation functions -- fusion probability plotted
against center-of-mass energy above the barrier --
as computed by both direct simulation and importance
sampling, with the same additional force as used
in Fig.\ref{fig:three}.
Each point was obtained using 1000 trajectories,
and the result is displayed with error bars, as
estimated from the numerical data.
The solid line represents an analytical formula
which closely approximates the fusion probability
over the region shown.\footnote{
This was obtained by running a very large number
of simulations for different values of 
energy above the barrier, and then fitting the
results to an exponential multiplied by a 
second-order polynomial.
The expected error associated with the curve
itself is everywhere smaller than the smallest
of the error bars shown in Fig.\ref{fig:five}.}
Again we see that, for approximately the same
computational effort, our importance sampling 
method gives significantly better results than
direct simulation.
For the point corresponding to 0.5 MeV above the
barrier, the error bar in Fig.\ref{fig:four} is
about 5.5 times bigger than that in Fig.\ref{fig:five}.
The efficiency gain of the importance sampling
approach is therefore about 30 ($\sim 5.5^2$, 
see Eq.\ref{eq:edv2}) in this case:
we would need to launch about $30\times 10^3$ trajectories
evolving under the original Langevin equation to
get the same degree of accuracy obtained in
Fig.\ref{fig:five} with $10^3$ trajectories.

The gain in efficiency becomes more dramatic when
we go to very small probabilities.
To show this we considered the reaction
$^{110}$Pd$+^{110}$Pd
($\mu\cong 0.174$, $X\cong 0.794$),
for which the extra push energy is 25.5 MeV.
Launching 250000 trajectories with an initial
center-of-mass energy of 1 MeV above the barrier,
we obtained a probability of fusion 
$P_{\rm fus} = (6.970\pm 0.268) \times 10^{-13}$.
This was computed using our importance sampling
method, with an additional force corresponding to 
$\Lambda=1.9\mu$; about 88\% of the trajectories evolving 
under the modified Langevin equation went to fusion.
Using Eq.\ref{eq:edv}, our result gives an 
efficiency gain of $E_{\Delta v}^G = 3.5 \times 10^9$!
We cannot compare our estimate of $P_{\rm fus}$ 
directly to an estimate obtained from simulating 
with the original Langevin equation, since we
would need to run $\sim 10^{12}$ trajectories
to have a decent chance of observing even a 
single fusion event.
Importance sampling is indispensible in this case:
we could not have calculated $P_{\rm fus}$ using direct 
simulations.

\section{An Exactly Solvable Model}
\label{sec:exact}

In addition to the numerical results of the
previous section, it is instructive
to consider an exactly solvable model.
Consider a particle in two dimensions which
falls at a uniform rate $v_z$ from a height
$h$, while experiencing random ``kicks'' in
the horizontal direction:
\begin{equation}
\dot z = -v_z \qquad,\qquad
\dot x = \hat\xi(t) \,\, ,
\end{equation}
where $\hat\xi(t)$ represents white noise corresponding
to a diffusion constant of unit magnitude:
$\langle \hat\xi(t) \hat\xi(t+s) \rangle = \delta(s)$.
We assume that the initial horizontal
location is zero, i.e.\
$x(0)=0$, and are interested in the
horizontal location of the particle
when it hits the ``ground'' ($z = 0$),
at time  $\tau = h/v_z$.
The motion in the $x$-direction is a 
Wiener process,
whose solution is a Gaussian distribution with a 
variance growing linearly with time.
Let us suppose we are interested in the
probability that the particle will
end to the right of the fixed point
$x_0$, i.e. $x(\tau)>x_0$.
Since the ensemble distribution of $x(\tau)$-values
is a Gaussian (with variance equal to $\tau$),
this probability is given in terms of an error function:
\begin{equation}
P[x(\tau)>x_0] \equiv F(x_0) =
{1 \over 2 }\Bigl[1-{\rm erf}(x_0/\sqrt{2\tau})\Bigr].
\end{equation}
For large values of $x_0$, this probability dies off
very rapidly, $P\sim e^{-x_0^2/2\tau}/x_0$,
therefore very many simulations would be needed to
compute $P$ to some desired relative accuracy $r$.

Let us now consider an additional force in the form
of a constant horizontal ``wind'', pushing the particle
in the direction of the point $x_0$.
The modified equations of motion are then
\begin{equation}
\dot z = -v_z \qquad,\qquad
\dot x = {\rm w } + \hat\xi(t) \,\, ,
\end{equation}
where w is the strength of the wind.
From Eq.\ref{eq:newda}, it follows that
$\Delta A = {\rm w} x(\tau) - {\rm w}^2\tau/2$.
From Eq.\ref{eq:ndv}, we can then compute the
minimum number of simulations necessary to obtain
$P$ to a relative accuracy $r$, using importance
sampling, for a particular value of wind strength:
\begin{equation}
N_{\rm w}^* = { 1 \over r^2 } \left\{ e^{{\rm w}^2 \tau } 
{ F(x_0+{\rm w}\tau) \over F^2(x_0) } - 1 \right\}.
\end{equation}
The efficiency gain is then:
\begin{equation}
\label{eq:egexact}
E_{\rm w}^G \equiv {N_0^* \over N_{\rm w}^*} = {{F(x_0)-F^2(x_0)}
\over{e^{{\rm w}^2 \tau}F(x_0+{\rm w}\tau )-F^2(x_0)}}.
\end{equation}

For the case $x_0=3.0$, $\tau=1.0$ ($P=1.35\times 10^{-3}$),
we have plotted efficiency gain as a function of
wind strength, in Fig.\ref{fig:six}.
We see that the optimal wind strength (at which we
obtain maximal efficiency) is ${\rm w}_{opt} = 3.157$. 
For this value of w, the number of trajectories
needed to estimate $P$ using
direct simulation of the original Wiener process, 
is about 220 times the number needed to estimate
$P$ (to the same degree of accuracy) with importance
sampling.

It is interesting to note that even for $x_0=0$
(for which half the trajectories fall to the right of 
$x_0$), we gain efficiency by using importance sampling.
In this case, for $\tau=1$, the maximal efficiency gain
(about 1.8) is achieved with a wind strength ${\rm w}=0.6125$, 
as shown in Fig.\ref{fig:seven}.

The maximal efficiency gain grows as the probability
$P$ becomes small.
Using an asymptotic expression for the error function,
we have
\begin{equation}
F(x_0) \rightarrow
\Biggl( {\tau\over 2\pi} \Biggr)^{1/2}
{1\over x_0}
\exp(-x_0^2/2\tau) \qquad (\tau\,\,{\rm fixed}\, ,\,x_0\rightarrow\infty),
\end{equation}
from which it follows that
\begin{equation}
N_w^* \rightarrow {x_0\over r^2}
\exp \Bigl[
(x_0-{\rm w}\tau)^2/2\tau \Bigr].
\end{equation}
This formula nicely encapsulates the dramatic
efficiency gain achieved for small values of $P$
(i.e.\ large $x_0$).
Without importance sampling, i.e.\ setting
w$=0$, the number of trajectories needed grows
exponentially in $x_0^2$:
$N_0^*\sim \exp(x_0^2/2\tau)$ (dominant contribution).
However, using importance sampling, with the optimal
wind value (${\rm w}_{opt} = x_0/\tau$), the number needed
grows linearly with $x_0$: $N_{\rm w}^* \sim x_0$. 
Thus, 
\begin{equation}
N^*_{\rm w} \sim \sqrt{ \ln N^*_0 }.
\end{equation}
Even when $N_0^*$ is tremendously large, $N_{\rm w}^*$
may remain modest (for ${\rm w}={\rm w}_{opt}$).
For instance, for $x_0=6$ and $\tau=1$ we have
$P[x(\tau)>x_0]=.99\times 10^{-9}$.
One would need $\sim 10^{12}$ direct simulations to
compute this probability to 10\% accuracy.
Using importance sampling, one can achieve the same
accuracy with fewer than 1000 trajectories.
Numerical results (with ${\rm w}=6.0$) gave us
$P[x(\tau)>6.0]=(1.04 \pm 0.10)\times 10^{-9}$
after only 700 trajectories.

\section{Conclusions}
\label{sec:conclusions}

In this paper we have developed a method for
computing the probabilities of rare events --
exemplified by the fusion of two nuclei --
for processes described by Langevin dynamics.
The method, based on the idea of importance sampling,
is straightforward to implement, quite general,
and can lead to a very large increase in
computational efficiency.
For these reasons we believe it represents a
very practical tool for using numerical simulations
to compute small probabilities.
Indeed, with our method, we were easily able to
estimate a fusion probability, within a schematic
model of nuclear collisions (see the end of
Section \ref{sec:numres}), that would have 
been essentially impossible to estimate from direct
simulations of the process in question.
We see every reason to expect similar results
when combining the method with more realistic
semiclassical models of nuclear dynamics.

\section*{Acknowledgments}

The authors wish to acknowledge stimulating
conversations with Janusz Skalski
and W\l adek \' Swi\c atecki.
This research grew out of conversations among
the authors at the XXV Mazurian Lakes School of
Physics (Piaski, Poland, 1997), and was continued
during reciprocal visits to the Institute for
Nuclear Studies in Poland, and Los Alamos National
Laboratory in the United States.
Financial support for this work came from the
Polish-American Maria Sk\l odowska-Curie Joint
Fund II, under project PAA/NSF-96-253.
J.B.\ also acknowledges support from the
Committee of Scientific Research of Poland
(KBN Grant No.\ 2-P03B-143-10).

\begin{figure}
\caption{
Thirty trajectories simulated using the schematic
model of nuclear collisions, Eq.\ref{eq:schem_orig}.
The system is $^{100}$Zr$+^{100}$Zr, at 0.8 MeV
above the interaction barrier.
Two trajectories lead to fusion; the rest to
reseparation.}
\label{fig:one}
\end{figure}

\begin{figure}
\caption{
The circle indicates the saddle point region of the 
potential energy, and the arrows show the direction
of the extra force chosen to push the system toward
fusion.  
Also shown are two (deterministic) trajectories,
one ending in fusion, the other in reseparation.}
\label{fig:two}
\end{figure}

\begin{figure}
\caption{
Convergence of the estimator 
$P_{\rm fus}^{(N)}$
with number of trajectories simulated ($N$), for both
direct simulation (solid line), and using 
importance sampling (dashed line).}
\label{fig:three}
\end{figure}

\begin{figure}
\caption{
Excitation function computed using direct simulation,
with 1000 trajectories for each point.
(The solid line is an analytical estimate extracted
from a much larger number of simulations.)}
\label{fig:four}
\end{figure}

\begin{figure}
\caption{
Same as Fig.\ref{fig:four}, but computed using
importance sampling instead of direct simulation.}
\label{fig:five}
\end{figure}

\begin{figure}
\caption{
Efficiency gain as a function of wind strength,
Eq.\ref{eq:egexact}, for $x_0=3$
($P=1.35\times 10^{-3}$).
At the optimal wind value, the gain is around
220.}
\label{fig:six}
\end{figure}

\begin{figure}
\caption{
Same as Fig.\ref{fig:six}, but for $x_0=0$.
Here, $P=0.5$, so there would be no problem
in estimating this probability from direct 
simulations.
Nevertheless, there is an efficiency gain of
nearly two, when using importance sampling.}
\label{fig:seven}
\end{figure}

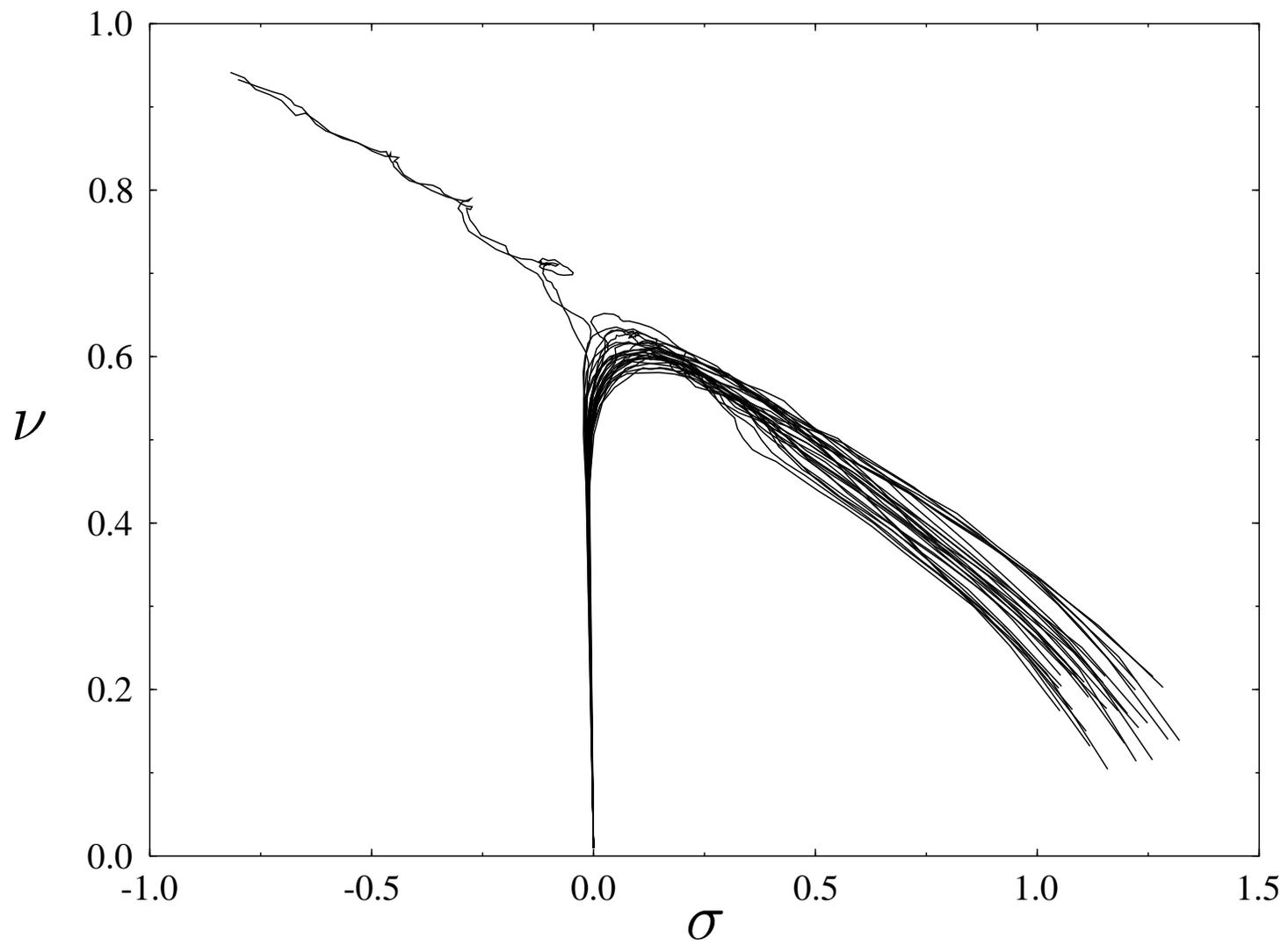
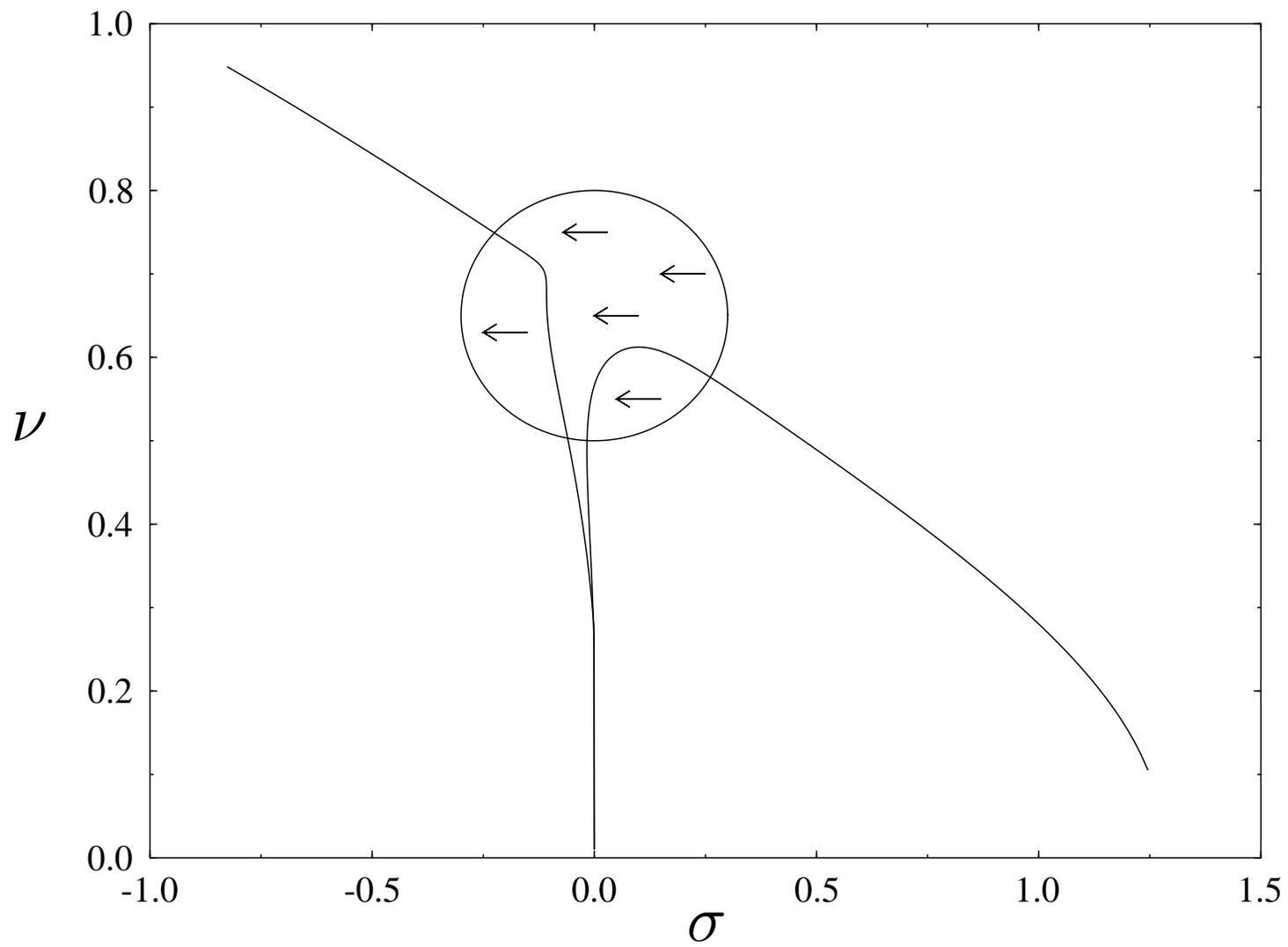
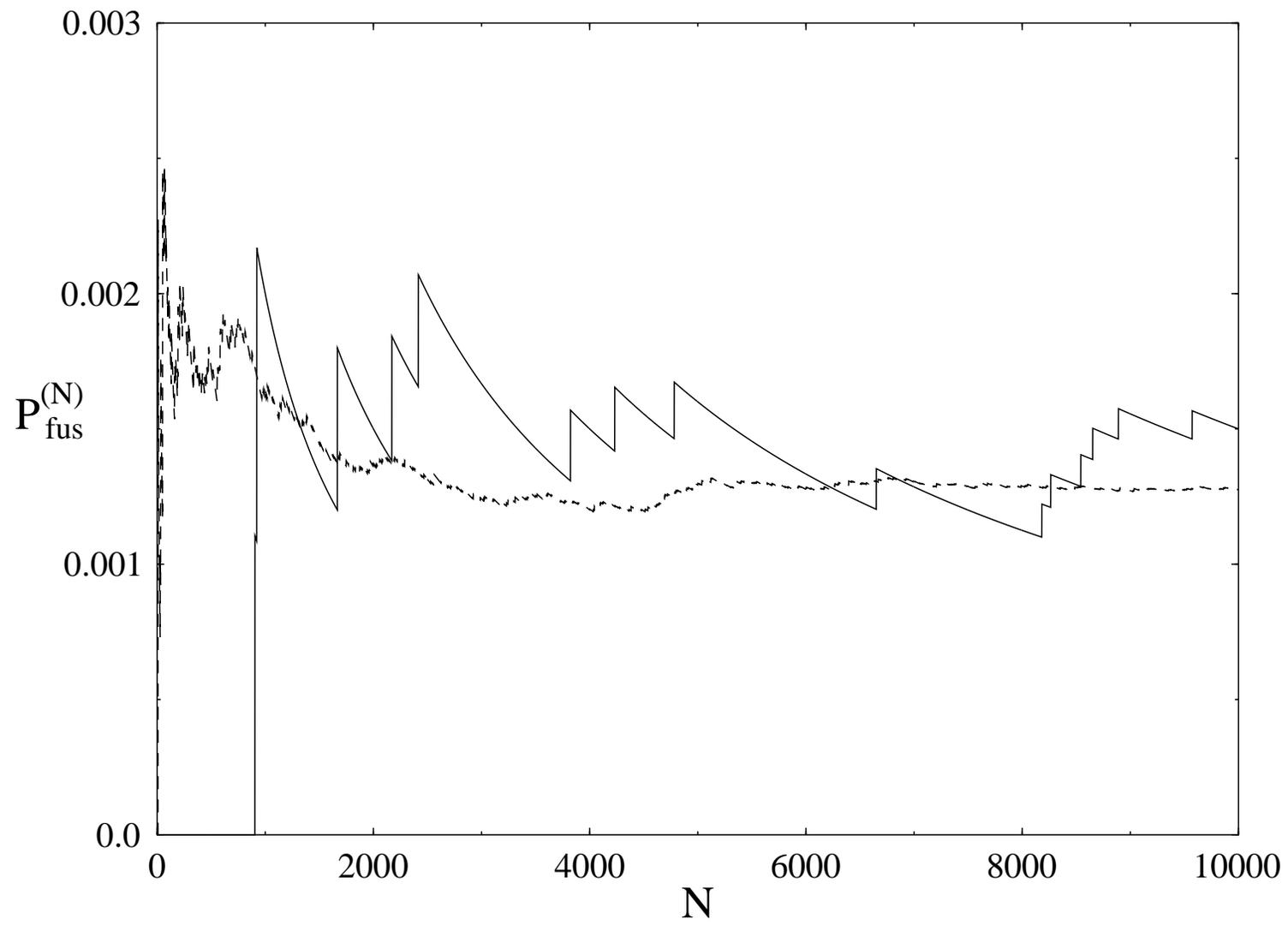
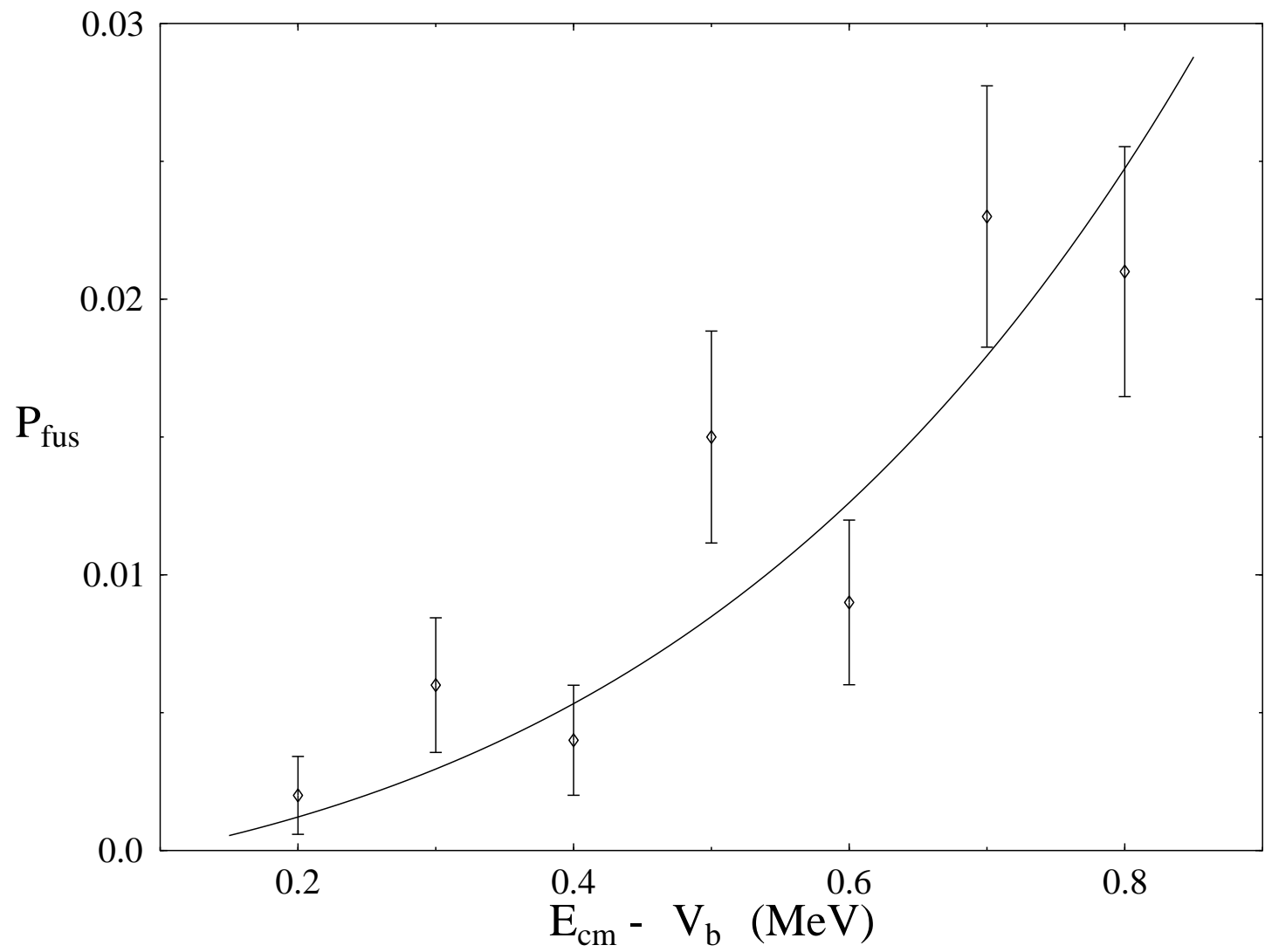
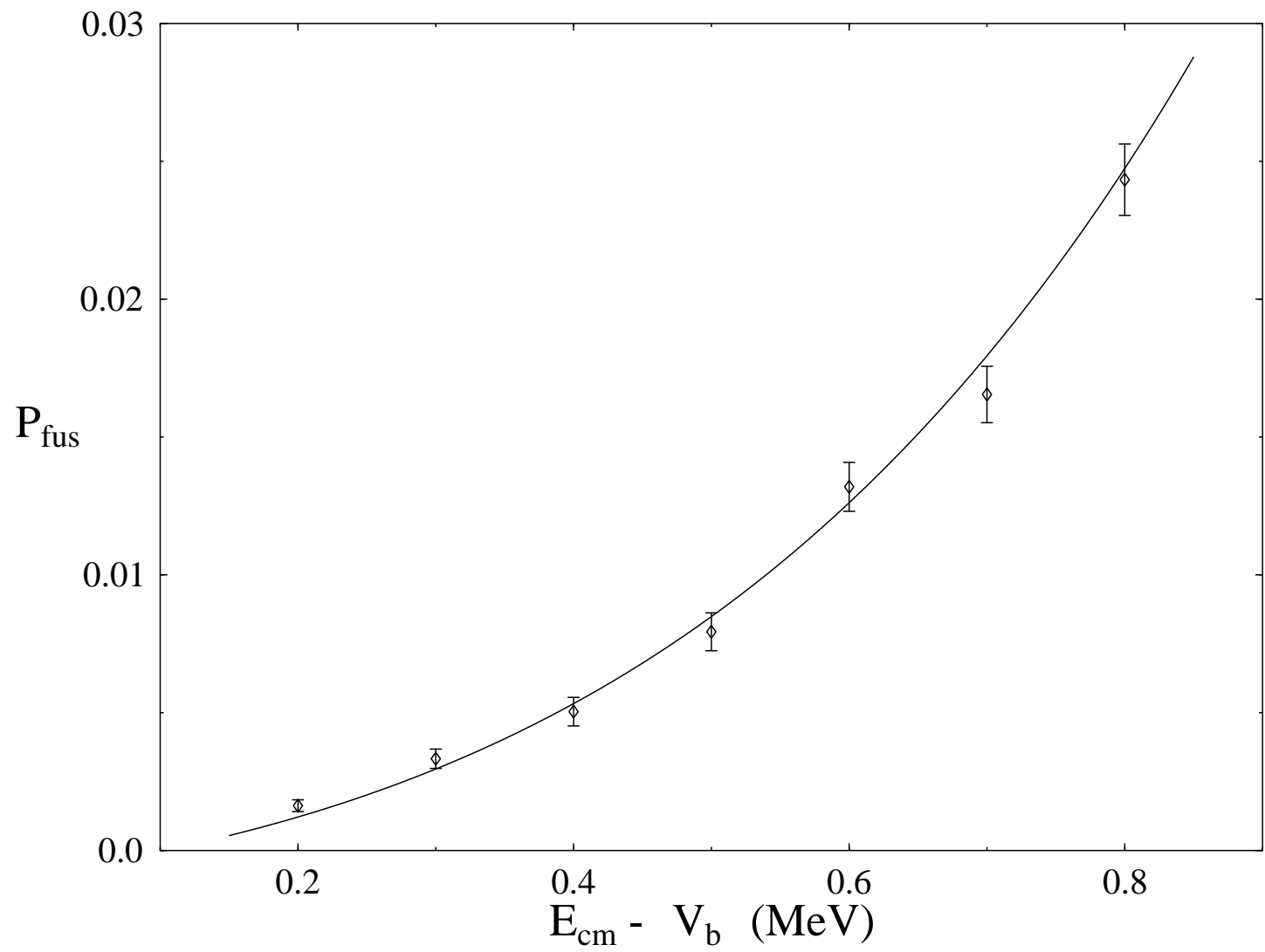
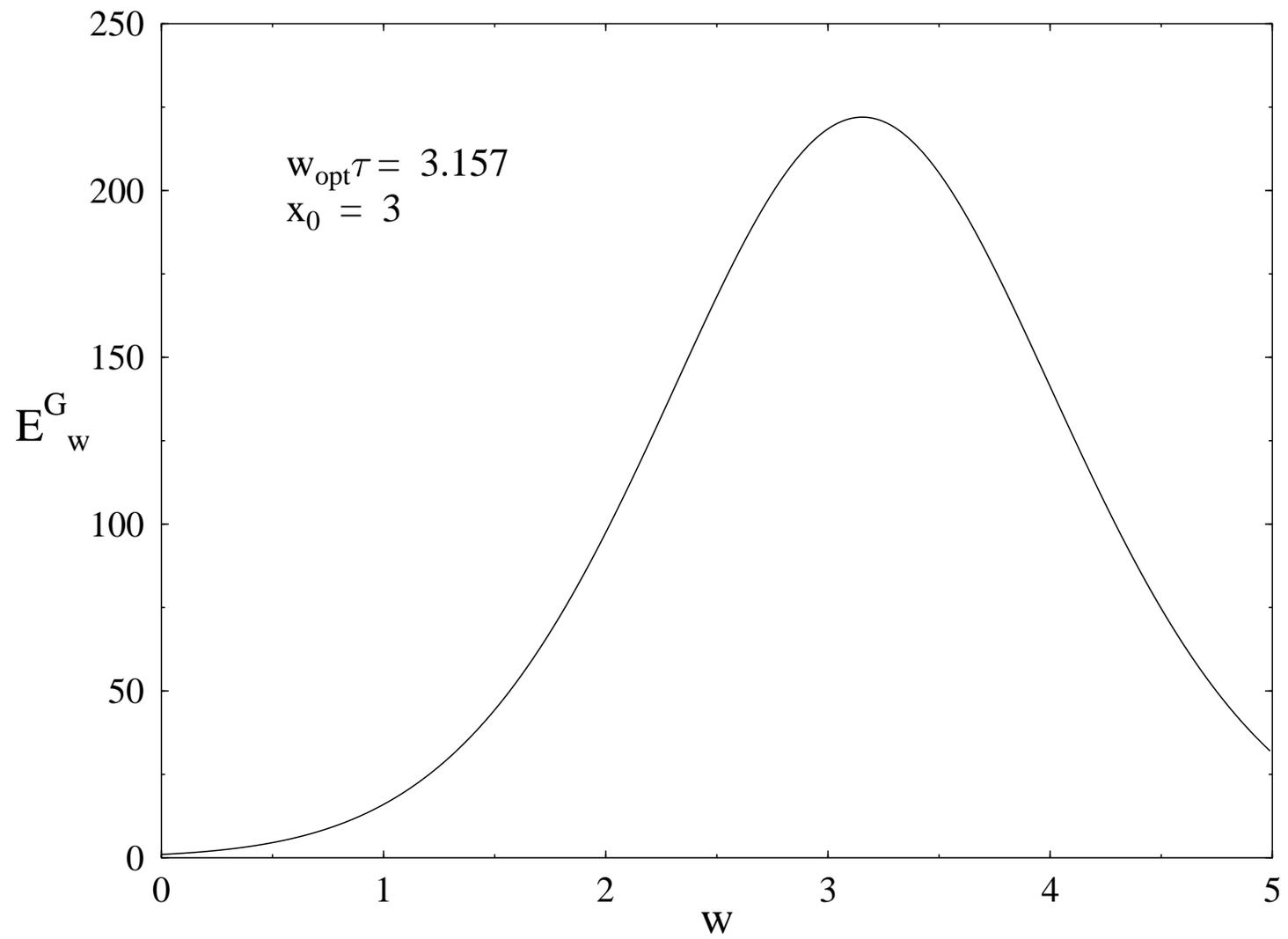
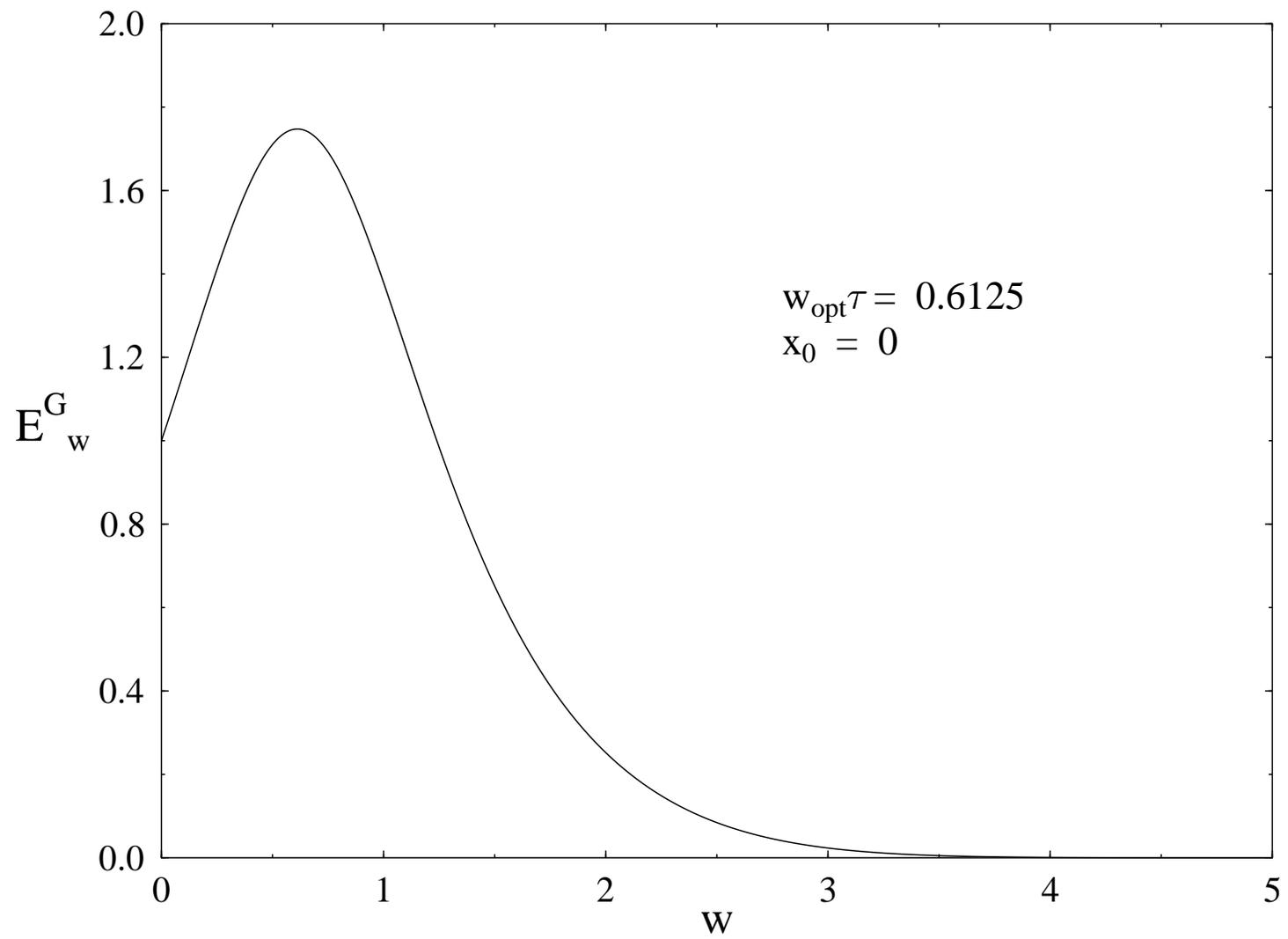
\end{document}